# Analysis of bacterial population growth using extended logistic Growth model with distributed delay

Tahani Ali Omer
Department of Mathematics and Statistics
University of Missouri-Kansas City
Kansas City, Missouri 64110-2499, USA.
tao34c@mail.umkc.edu

## Abstract

In the present work, we develop a delayed Logistic growth model to study the effects of decontamination on the bacterial population in the ambient environment. Using the linear stability analysis, we study different case scenarios, where bacterial population may establish at the positive equilibrium or go extinct due to increased decontamination. The results are verified using numerical simulation of the model.

## INTRODUCTION

The Logistic model was developed by Belgian mathematician Pierre Verhulst (1838), who suggested that the rate of population increase maybe limited. The dynamics of the population was described by the differential equation:

$$\frac{dN(t)}{dt} = rN(t)\left[1 - \frac{N(t)}{K}\right], \quad (1)$$

where N (t) is the population size at time t, r is the intrinsic growth rate, and K (>0) is the carrying capacity of the population.

Logistic model combines two ecological processes: reproduction (growth) and competition (self-limiting). Both processes depend on population numbers (or density). The rate of both processes match to the mass-action low with coefficients: r for reproduction and r/K for competition.

The Logistic model provides a good mathematical description for many biological populations of microorganisms, plants, and animals. It is also widely used in statistics, economics, medicine [1] [2], physics, chemistry, and other applications.

The Logistic model contains some deficiencies. It does not account for at least some individual mortality that can occur at any time, including during the exponential growth stage. This difficulty can be partly overcome by considering N (t) in the equation as a number that represents the population's net growth [3]. If so, then any mortality that might occur during the 'lag', 'exponential' and 'stationary' stages is reflected in the observed magnitude of the model's parameters. In particular it has been suggested. (e.g., Corradini & Peleg, 2006) that the momentary growth rate might be proportional to $N(t)^a$ and to $\left[1 - \frac{N(t)}{K}\right]^b$, in which case the logistic equation will become:

$$\frac{dN(t)}{dt} = rN(t)^a \left[1 - \frac{N(t)}{K}\right]^b, \quad (2)$$

where a is the growth regulator and b is the depletion factor.

The Logistic model assumed that the growth rate of a population at any time t depends on the relative number of individuals at that time. In practice, the process of reproduction is not instantaneous [4]. Therefore, one way of improving the Logistic growth model Eq. (1) is the following more realistic logistic equation:

$$\frac{dN(t)}{dt} = rN(t) \left[1 - \frac{N(t-\tau)}{K}\right], \quad (3)$$

Eq. (3) proposed by Hutchinson (1948) [5]. Where r and K have the same meaning as in the Logistic equation Eq. (1), $\tau$ >0 is the discrete delay term.

Eq. (3) means that the controlling effect depends on the population at a fixed earlier time $(t - \tau)$, rather than the present time t, the effects of time delay t>0 on stability of equilibria has been well-studied (Kaung 1999). In a more realistic model the delay effect should be an average over past populations. This results in an equation with a distributed delay. Thus, another way of improving the Logistic growth model Eq. (1) was to include a delay term to examine a cumulative effect in the death rate of a species, depending on the population at all times from the start of the experiment. The model is an integro-differential equation:

$$\frac{dN(t)}{dt} = rN(t)\left[1 - \frac{1}{K}\int_{-\infty}^{t} N(s)\,G(t-s)\,ds\right], \quad (4)$$

Eq. (4) was initially proposed by Volterra (1934) [6] [7], where r and K have the same meaning as in the logistic equation Eq. (1), G(t) is a weighting factor which indicates how much emphasis should be given to the size of the population at earlier times to determine the present effect on resource availability. Eq. (4) it includes a type of delay that is known as distributed delay which is the sum of infinitely numerous small delays in the form of an integral.

Also, to further improve the Logistic growth model Eq. (1) MacDonald (1978) [8] [9] discussed the following integro-differential equation:

$$\frac{dN(t)}{dt} = rN(t)\left[1 - \frac{N(t)}{K} - \int_{0}^{t} N(s)\,G(t-s)\,ds\right], \quad (5)$$

This includes both distributed delay integral term and the traditional $\frac{N(t)}{K}$ term. Here r and k are positive, and instantaneous self-crowding term G (t) is accompanied by a population term N (t).

In the present work we develop a more realistic model which takes advantage of the above-mentioned modeling approaches and we employ the model to investigate the effects of decontamination on the bacterial population [10] [11].

## The Model

Our goal is to use previous modeling approaches along with knowledge on how bacterial grow and develop a model describing the bacterial growth that is affected by decontamination.

Bacterial population growth cells reproduce by dividing into two cells. Each individual (bacterium, virus, or microbe) takes about the same time to mature and divide. This principle is commonly used with the Peleg's model Eq. (2) and MacDonald's model Eq. (5).

Peleg's model Eq. (2) is based on the replacement of r (the growth rate) in Eq. (1) by a term that presumably accounts for the initial population and links the 'lag time', and 'maximum specific growth rate' in a predetermined relationship. Moreover, it is based on assuming the momentary growth rate might be proportional to $N(t)^a$ and to $\left[1 - \frac{N(t)}{K}\right]^b$, where a is the growth regulator and b is the depletion factor, in order to overcome the difficulty that the logistic model does not account for at least some cell mortality that can occur at any time, including during the exponential growth stage.

Also, we will be using MacDonald's model Eq. (5) with a type of delay known as distributed delay which is the sum of infinitely numerous small delays in the form of an integral. The model below uses both previous models and adds a term with a decontamination rate. To develop a mathematical model of bacterial population growth using extended logistic growth model with distributed delay

$$\frac{dN(t)}{dt} = rN(t)^a \left[1 - \frac{N(t)}{K} - \int_0^t N(s)\, G(t-s)\, ds\right]^b - d\, N(t), \quad (6)$$

Following is the table of variables and parameters:

| Symbols | Descriptions |
| --- | --- |
| N(t) | The bacterial population size at time t |
| r | The growth rate of bacteria |
| K | The carrying capacity |
| a | Growth regulator |
| b | Depletion factor |
| G(t-s) | The weight functions. |
| d | The decontamination rate of bacteria. |
| S | Delay term. |

# Analysis at the reduced model

Our model Eq. (6) cannot be integrated analytically, therefore we will focus on the qualitative behavior of the solution.

First, we will analyze at the reduced model

$$\frac{dN(t)}{dt} = rN(t)^a \left[1 - \frac{N(t)}{K}\right]^b - dN, \quad (7)$$

which is the reduced form of Eq. (6) under the assumption that there is no delay effect (i.e., G (t) = 0 for all t∈R). We consider two cases.

### Analysis at the reduced model Case 1: a = b = 1.

This will reduce the model Eq. (7) to the original logistic growth model Eq. (1) with an additional decontamination term. In this case to determine the equilibrium solution we know the rate of change of the population will be equal to zero, so there are two equilibria: N* = 0 And N* = K (1 – d/r). Moreover, if we let $f(x) = rN(t)\left[1 - \frac{N(t)}{K}\right] - dN$, then asymptotic stability is determined by the sign of the derivative of f(x) evaluated at the equilibrium point. For example:

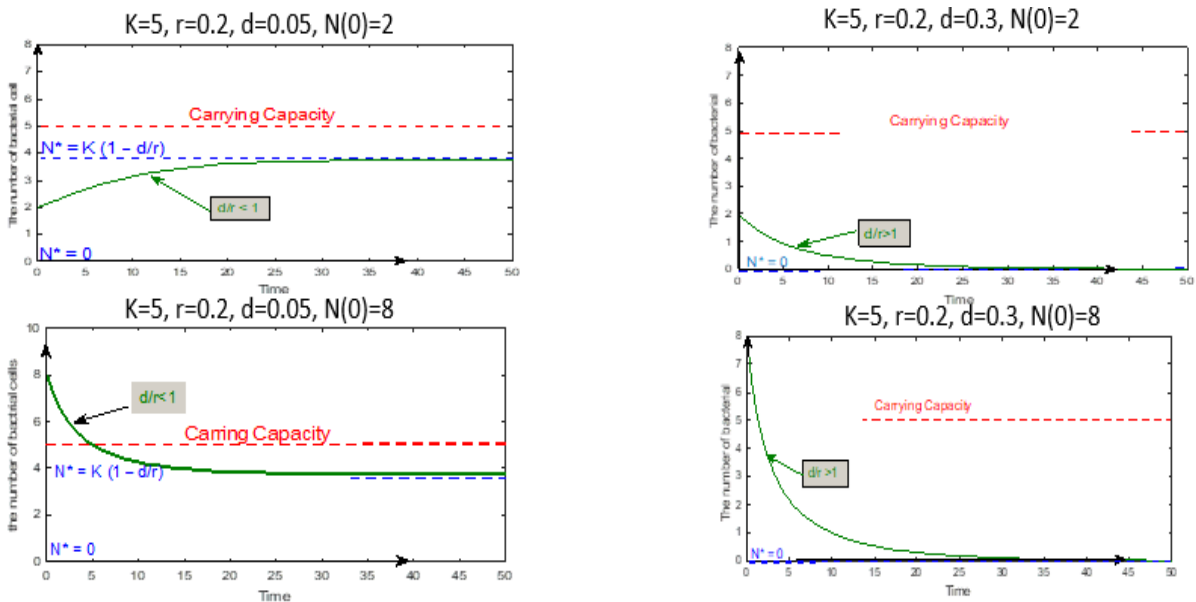

**Fig. 1:** Numerical simulations of model (7). When a = b =1, $\frac{d}{r} < 1$, and $\frac{d}{r} > 1$.

From the Fig.1 the solution converges to zero, otherwise the solution reaches to the positive equilibrium.

For the biological interpretation we will summarize the equilibrium solution and their asymptotic stability with respect to the ratio d/r in the following table:

| Equilibrium. | Condition. | Outcome. | Biological interpretation |
| --- | --- | --- | --- |
| N* = 0 | d/r > 1 | Asymptotically stable | The bacterial population will go extinct. |
| N* = 0 | d/r < 1 | unstable | The bacterial population will establish |
| N* = K (1 – d/r) | d/r > 1 | impossible | No biological meaning |
| N* = K (1 – d/r) | d/r < 1 | Asymptotically stable | The bacterial population will go extinct. |

Analysis at the reduced model Case 2:  $a \geq 0, \ b \geq 0$

In this case the rate of change of the population equal to zero.

$$rN(t)^a \left[1 - \frac{N(t)}{K}\right]^b - dN = 0$$

Imply, $dN = rN(t)^a \left[1 - \frac{N(t)}{K}\right]^b$

As illustrated in Fig.2 the intersections of the line dN and the function b (N) = $rN(t)^a \left[1 - \frac{N(t)}{K}\right]^b$ are the equilibrium solution of the Model.

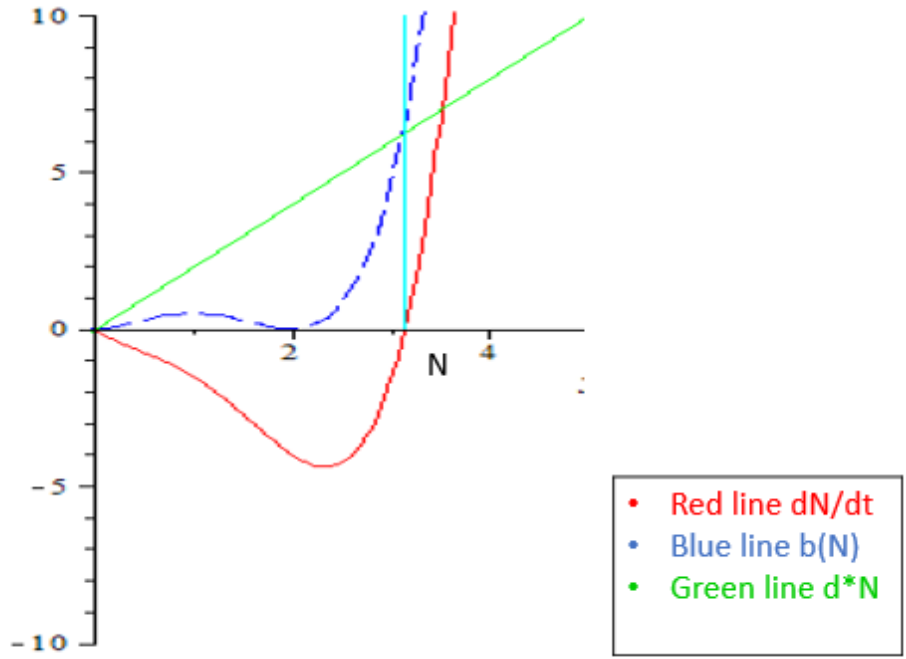

Fig. 2: Numerical simulations of model (7). When $a \geq 0 \; and \; b \geq 0$

Moreover, if we let $b(N) = rN(t)^a \left[1 - \frac{N(t)}{K}\right]^b$, and N* be an equilibrium solution of the model. Then by the derivative test the sign of b' (N*)-d determines the stability for N*. But note that b' (N*) is the slope of the b (N) at N* and d is the slope of the dN everywhere including N*. Hence, the difference between the slope of b (N) at each intersection point and the decontamination rate d determines the asymptotic stability of the equilibrium solution.

For additional analysis we will only consider four selected combinations of powers a and b which are (1) a > 1 and b odd; (2) a> 1 and b even; (3) $a \leq 1$ and b odd; (4) $a \geq 1$ and b even, those possible cases have been summarized in Fig.3, where survival or extinction-survival may occur based on the values of a & b.

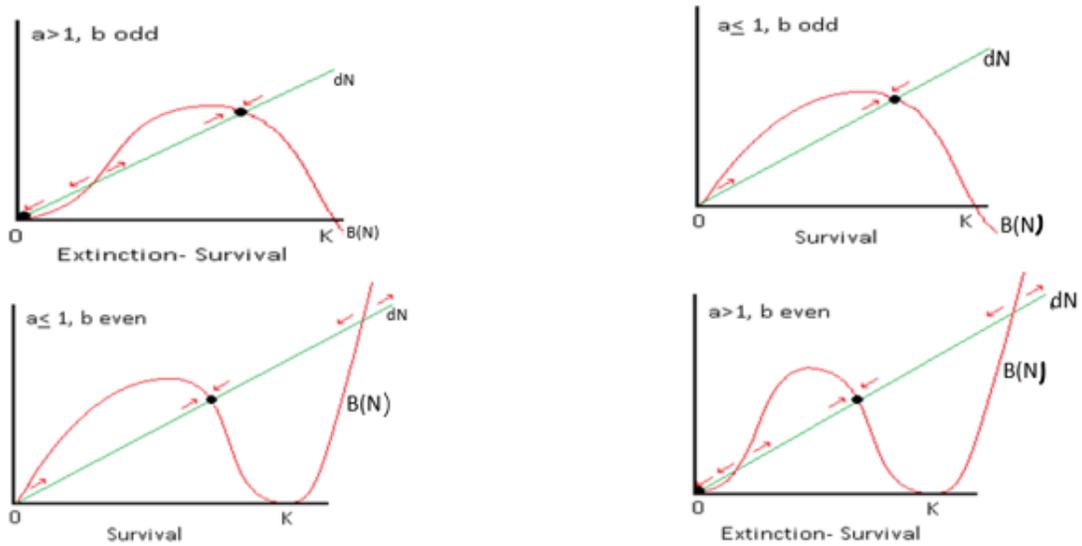

**Fig. 3:** Numerical simulations of model (7). When a > 1 and b odd;
a > 1 and b even; $a \leq 1$ and b odd; $a \geq 1$ and b even.

For the biological interpretation we will summarize the equilibrium solution and their asymptotic stability in the following table:

| Condition | Maximum Number Of Equilibrium | Outcome | Biological Interpretation |
|---|---|---|---|
| a > 1<br>b odd | Three | 1. Stable.<br>2. Unstable.<br>3. Stable | The bacterial population will go extinct or survival depending on population initial number. |
| a ≤ 1<br>b odd | Two | 1. Unstable.<br>2. Stable. | The bacterial population will survival. |
| a ≤ 1<br>b even | Three | 1. Unstable<br>2. Stable.<br>3. Unstable. | The bacterial population will survival. |
| a > 1<br>b even | Four | 1. Stable.<br>2. Unstable.<br>3. Stable.<br>4. Unstable | The bacterial population will go extinct or survival depending on population initial number. |

## Conclusion

It is our goal to analyze the general model, using the same method provided in this paper. In a conclusion, analysis of delayed Logistic growth models can provide a mathematical framework for a better understanding of bacterial growth and survival affected by different decontamination policies.

## ACKNOWLEDGEMENT

The research on "**Analysis of bacterial population growth using extended logistic growth model with distributed delay**" was given to me when I was a research assistant at the University of Missouri, Kansas City. I completed this paper under the supervision of Dr. Bani-Yaghoub. I would like to express my gratitude to him for giving me productive advice, for being a source of motivation, and for providing me with materials I could not have discovered on my own. Thank you, Dr. Bani! Words can never be enough to thank your kindness.